\documentclass{article}

\usepackage[final,dvips]{graphicx}      % lets me use postscript graphs
\usepackage{feynmp}                     % feynman diagram generation
\usepackage{amssymb}                    % extra amstex symbols

\skip\footins 39pt plus 4pt minus 2pt
\def\footnoterule{\kern-19pt\hrule width.5in\kern18.6pt}

\newcommand{\eq}[1]{(\ref{#1})}

\newcommand{\fig}[1]{Figure~\ref{#1}}
\newcommand{\Fig}[1]{Figure~\ref{#1}}

\newcommand{\ignore}[1]{{}}

\newcommand{\be}{\begin{equation}}
\newcommand{\ee}{\end{equation}}

\newcommand{\bea}{\begin{eqnarray}}
\newcommand{\eea}{\end{eqnarray}}

\begin{document}

\renewcommand{\topfraction}{0.99}
\renewcommand{\bottomfraction}{0.99}
\twocolumn[\hsize\textwidth\columnwidth\hsize\csname
@twocolumnfalse\endcsname

\begin{center}
{\Large \bf Fermions, Gauge Theories, and the Sinc Function
  Representation for Feynman Diagrams }  \\
\bigskip
{\large Dmitri Petrov$^{1}$, Richard Easther$^{1,2}$, Gerald
  Guralnik$^{1}$, Stephen Hahn$^{1}$ and Wei-Mun Wang$^{3}$} \\
\smallskip
{\small $^{1}$ Department of Physics,  Brown University,
  Providence, RI  02912, USA. \\
 $^{2}$ Department of Physics, Columbia University, New York, NY
 10027, USA \\
 $^{3}$ Department of Finance, The Wharton School, University of Pennsylvania,  Philadelphia, PA 19104, USA.  }
\medskip

\begin{abstract}
  
  We extend our new approach for numeric evaluation of Feynman diagrams to
  integrals that include fermionic and vector propagators. In this
  initial discussion we begin by deriving the Sinc function
  representation for the propagators of spin-$1\over2$ and spin-$1$
  fields and exploring their properties.  We show that the attributes
  of the spin-$0$ propagator which allowed us to derive the Sinc
  function representation for scalar field Feynman integrals are
  shared by fields with non-zero spin. We then investigate the application
  of the Sinc function representation to simple QED diagrams,
  including first order corrections to the propagators and the vertex.

\end{abstract}

\end{center}

BROWN-HET-1125

\bigskip
\mbox{}
]

%%%%%%%%%%%%%%%%%%%%%%%%%%%%%%%%%%%%%%%%%%%%%%%%%%%%%%%%%%%%%%%%%%%%%%%%

\section{Introduction}

In conjunction with several colleagues, four of the present
authors are in the process of developing a novel framework for
non-perturbative numerical calculations in Quantum Field Theory. This
{\it Source Galerkin Method\/}
\cite{GarciaET1994a,GarciaET1996a,LawsonET1996a,Hahn1998a,EmirdagET1999a}
is free of many of the problems that plague Monte Carlo calculations,
in part because it works on the continuum and treats bosons and
fermions in a largely symmetrical manner.  A useful and surprising
by-product of this ``exact'' method is that some of the computational
techniques we developed in order to implement it are also applicable
to the numerical evaluation of Feynman diagrams in conventional
perturbation theory.

The resulting new method of numerically evaluating Feynman diagrams,
called the Sinc function representation, was introduced by three of us
in \cite{EastherET1999c}, and
extensively tested in \cite{EastherET1999e}.  It relies on the
observation that almost any integral can be approximated as an
infinite sum, via the use of generalized Sinc functions.\footnote{We
  use the capital S in Sinc to differentiate these functions from the
  standard form, $\mbox{sinc}(x) = \sin{x}/x$.}  Using this
representation, we develop approximate versions of the field
propagators where the spatial (or momentum) dependence only appears in
terms like $\exp(-x^2)$, so that all the integrals over vertex
locations or internal momenta are reduced to Gaussian integrals, which
can be performed analytically. After these integrals are evaluated, the
diagrams are represented as a multi-dimensional but rapidly convergent
sum, which is evaluated numerically.

The initial investigation of the Sinc function representation of the
propagators of fields with non-zero spin \cite{Hahn1998a} and their
application to Feynman diagram calculations \cite{Wang1998a} strongly
suggested that the Sinc function representation would be useful for
the calculation of diagrams in gauge theories. However, until now our
systematic development of the Sinc function representation has been
limited to scalar fields, which allowed us to develop our procedures
in a relatively simple, index-free environment.  Here we
demonstrate that our evaluation methods generalize directly to fields
with spin, and thus provide a powerful tool for calculations in more
general theories. In this paper we focus on diagrams whose analytic
properties are well understood, and calculate a variety of second
order QED diagrams. We will extend our calculations to more
complicated graphs in future work. We argue that these calculations---in
conjunction with the higher order evaluations we have performed for
scalar diagrams \cite{EastherET1999e}---provide strong evidence that the
Sinc function representation will be a useful calculational tool
for fermionic and gauge theories.

In the next section we recapitulate
the most important facts about the generalized Sinc function and in
the following section we derive the approximate versions of vector and
Dirac field propagators, checking their accuracy by comparing them to
the exact forms. In Section 4 we explicitly evaluate several diagrams,
showing that we can reproduce the results of conventional second order
perturbation theory to very high accuracy while using insignificant
amounts of computer time.

In future papers we will calculate higher order diagrams while
providing a development of renormalization theory that will allow a
straightforward automation of the calculation of perturbative
quantities. We believe that the Sinc function representation
will be useful in a wide range of applications from the evaluation of
experimental results to the easy prototyping of model theories.

\section{Sinc Functions}

An overview of the properties of Sinc functions relevant to Feynman
diagram calculations was given in \cite{EastherET1999c}, and
\cite{StengerBK1,Higgins1985a} give more thorough discussions. Here we
simply restate a few important results.  A generalized Sinc function
is defined by
\be
S_k(h,x) = \frac{\sin{\left[\pi (x-kh)/h\right]}}{\pi(x-kh)/h}.
\ee
Using an integral representation of this function, it is easy to show
that any function $f(z)$ which is analytical on a strip of a complex
plane centered around the real axis and with width $2d$, has the following
approximation
\be
\label{sinc_exp}
\int\limits_{-\infty}^{\infty}f(z)dz\approx\sum_{k=-\infty}^{\infty}\int
\limits_{-\infty}^{+\infty}dzf(kh)S_k(h,z)=h\sum_{k=-\infty}^{\infty}f(kh).
\ee
The error of this approximation
\be
\Delta
f(h)=\int\limits^{+\infty}_{-\infty}dzf(z)-h\sum^{+\infty}_{k=-\infty}
f(kh) 
\ee
satisfies 
\be
|\Delta f(h)|\le C\frac{e^{-\pi d/h}}{2\sinh(\pi d/h)}.
\ee
Here $C$ is a number independent of $h$. This formula shows that if
$h<d$, the difference between the exact and the approximate results
decreases exponentially with $h$. 

When we calculate the Sinc function representation there are two
sources of error. Firstly, the quantity we are calculating differs
from the exact value by an amount that depends on $h$, and secondly we
are performing a numerical evaluation of infinite sum, and doing so
inevitably introduces truncation errors and inaccuracies due to the
finite precision of computer arithmetic.  Consequently, by choosing
$h$ we attempt to establish the amount of accuracy we hope to obtain
from our calculation, but we must ensure that our numerical evaluation
of the resulting sum is carried out to a commensurate level of
precision.

\section{Propagators}
\subsection{Coordinate Space}

In this section we obtain Sinc function representations of the 
fermion and vector propagators in coordinate and momentum
space.\footnote{In this paper we only directly consider massive, spin
  $1\over2$ fermions, but will often ignore this qualification when
  referring to the fermion propagator. However, these calculations
  will generalize to massless spin-$1\over2$ fields and spin-$3/2$
  fields. Likewise we will use  ``photon'' interchangeably with
  ``vector'' since our explicit examples are drawn from QED, although we
  can compute diagrams containing massive gauge fields.}
In order to achieve this, an approach similar to the one used in
\cite{EastherET1999c} will be applied to calculate coordinate space
propagators. The momentum space propagators will be obtained by
Fourier transformation.

We use the fermion propagator to illustrate this calculation.  Almost
identical procedures are used to obtain expressions for the other
propagators.

In Euclidean space, the fermion propagator can be written as
\be
S_F(x)=\int\frac{d^4_Ep}{(2\pi)^4}\frac{-\gamma^{\mu}p_{\mu}+m}{p^2+m^2}
e^{ipx}.
\ee
After introducing a cut-off and exponentiating the denominator we obtain:
\begin{eqnarray}
S_F(x,\Lambda)&=&\int \limits^{\infty}_0ds\:e^{-sm^2}
    \int\frac{d^4_Ep}{(2\pi)^4}(m-\gamma^{\mu}p_{\mu})\times\nonumber\\ 
&& \exp{\left(-\left(s+\frac{1}{\Lambda^2}\right)p^2+ipx\right)}
\label{cutoff}.
\end{eqnarray}
The second integral in (\ref{cutoff}) is a Gaussian, which can be
easily computed analytically:
\bea
S_F(x,\Lambda)&=&\int
\limits^{\infty}_0ds\:e^{-sm^2}\frac{1}{32\pi^2\left
    (s+\frac{1}{\Lambda^2}\right )^3}\times\nonumber\\ 
&& \quad
\left [2m\left (s+\frac{1}{\Lambda^2}\right
  )-i\gamma^{\mu}x_{\mu}\right] \times \nonumber\\
&& \quad
\exp{\left (-\frac{x^2}{4\left (s+\frac{1}{\Lambda^2}\right )}\right )}.
\eea

To make $s$ dimensionless we shift $s$ to $s/M^2$, where $M$
is an arbitrary constant with dimensions of mass. The final result
does not depend on a specific value of $M$. After performing this
transformation and setting $s$ to be equal to $e^z$ we obtain:
\bea
S_F(x,\Lambda)&=&\int
\limits^{\infty}_{-\infty}dz\frac{M^2}{32\pi^2}\frac{\exp\left
    (z-e^z\frac{m^2}{M^2}\right )} 
{\left (s+\frac{M^2}{\Lambda^2}\right )^3}\times\nonumber\\
&& \quad
\left [ 2m\left (s+\frac{M^2}{\Lambda^2}\right
  )-iM^2\gamma^{\mu}x_{\mu} \right ] \times\nonumber\\
&& \quad
\exp{\left(\frac{x^2M^2} 
{4\left(s+\frac{M^2}{\Lambda^2} \right ) }\right )}. 
\eea
At this point we use the Sinc expansion (\ref{sinc_exp}) to arrive at
the final form of Sinc function approximation to the fermion
propagator in coordinate space.
\bea
S_{Fh}(x,\Lambda)=\frac{hM^2}{32\pi^2}\sum\limits^{+\infty}_{k_1=-\infty}
\frac{e^{k_1h}}{\left (e^{k_1h}+\frac{M^2}{\Lambda^2}\right
  )^3}\times\nonumber\\ 
\left [ 2m\left (e^{k_1h}+\frac{M^2}{\Lambda^2}\right ) -
  iM^2\gamma^\mu x_\mu\right ]\times\nonumber\\ 
\exp{\left (
    -e^{k_1h}\frac{m^2}{M^2}-\frac{x^2M^2}{4\left(e^{k_1h}+\frac{M^2}
        {\Lambda^2}\right)}\right)}. 
\label{ferm_sinc_coord}
\eea
In the above expression, and in the rest of this paper, we will use
the subscript $h$ to differentiate between the exact propagator and
its Sinc function approximation.  We also introduce a shorthand
notation
\be
\alpha(k_1,k_2,k_3,...)=e^{k_1h}+e^{k_2h}+e^{k_3h}+...+\frac{M^2}{\Lambda^2},
\ee
which makes algebraic expressions similar to (\ref{ferm_sinc_coord})
more compact.

By repeating all of the above steps, we find an approximation for the
scalar field propagator:
\bea
\label{sc_sinc_coord}
G_h(x,\Lambda)&=&\frac{M^2h}{(4\pi)^2}\sum\limits^{+\infty}_{k_1=-\infty}
  \frac{e^{k_1h}}{(\alpha(k_1))^2}\times\nonumber\\
&& \quad \exp{\left (-e^{k_1h}\frac{m^2}{M^2}-\frac{M^2x^2}{4\alpha(k_1)} \right )}
\eea
This form of scalar propagator becomes identical to the form derived
in \cite{EastherET1999c}, if $M$ is set equal to $m$. In
general, the numerical value of the parameter $M$ does not have any
significance for either scalar or fermion field propagators.

Previously, we avoided introducing undetermined parameters in the
theory by shifting $s$ to $s/m^2$ instead of $s/M^2$.
However, in the case of the photon, $m=0$, so we cannot set $M$ equal
to the particle mass without dividing by zero. Moreover, in order to
regulate infrared divergences we will find it is necessary to assign a
small mass, $m_\nu$, to the photon and then take the limit where this
mass goes to zero. In this case shifting $s$ to $s/m_\nu^2$
would produce terms involving $m_\nu/\Lambda$, where
$m_\nu\rightarrow 0$ and $\Lambda\rightarrow\infty$. We prefer not do
this, since this choice prevents us from handling ultraviolet and infrared
divergences independently.

In an arbitrary gauge, the Euclidean space photon propagator is
\bea
D^{\mu\nu}(x)&=&-\int\frac{d^4_Ep}{(2\pi)^4}\frac{1}{p^2+m_\nu^2} 
  \frac{1}{p^2} \times\nonumber\\
&&\quad (-\delta^{\mu\nu}p^2+(1-\xi)p^\mu p^\nu)e^{ipx}.
\label{phot_init}
\eea
There are two factors containing $p^2$ in the denominator of
\eq{phot_init}. The Sinc expanded form of this expression requires
independent exponentiation of each of these factors and consequently
there are two independent summation parameters, $k_1$ and $k_2$, in
the expression for the propagator:
\bea
D^{\mu\nu}_h(x,\Lambda)&=&-\frac{M^2h^2}{(4\pi)^2}\sum\limits_{k_1,k_2} 
  \frac{e^{k_1h}e^{k_2h}}{\alpha(k_1,k_2)}\times\nonumber\\
 && \exp{\left( -\frac{x^2M^2}{4\alpha(k_1,k_2)}-e^{k_2h}\frac{m_v^2}{M^2}
  \right)}\times\nonumber\\ 
&&\left[
  -\frac{1}{2}\delta^{\mu\nu} (3+\xi) \right. + \nonumber\\
&& \qquad \left. \frac{M^2}{4\alpha(k_1,k_2)}
  (\delta^{\mu\nu}x^2-x^\mu x^\nu)\right]. 
\label{phot_sinc_gen}
\eea

The double summation in \eq{phot_sinc_gen} makes the task of
evaluating Feynman diagrams containing many photon lines significantly
more difficult.  Fortunately it is possible to solve this problem by
choosing a particular gauge. In the Feynman gauge $(\xi=1)$ the photon
propagator is
\be
D^{\mu\nu}(x)=\int\frac{d^4_Ep}{(2\pi)^4}\frac{\delta^{\mu\nu}} 
  {p^2+m_\nu^2}e^{ipx}.
\ee
This can be approximated by
\bea
D^{\mu\nu}_h(x,\Lambda)&=&\delta^{\mu\nu}\frac{M^2h}{(4\pi)^2} 
  \sum\limits_{k_1=-\infty}^{+\infty}\frac{e^{k_1h}} 
{(\alpha(k_1))^2}\times\nonumber\\
&& \quad
\exp{\left(-e^{k_1h}\frac{m_\nu^2}{M^2}-\frac{M^2x^2}{4\alpha(k_1)}
  \right )}. 
\label{phot_sinc_simpl}
\eea

At this point we have completed the derivation of Sinc expanded
versions of the propagators in coordinate space. However, for a number
of reasons it is more convenient to evaluate diagrams in momentum
space. The transition to momentum space significantly simplifies the
algebraic form of the Sinc function representations of the propagators.

\subsection{Momentum Space}

One way of obtaining approximate versions of the propagators in
momentum space is to take Fourier transformations of the corresponding
expressions in coordinate space. Alternatively we can start by Sinc
expanding exact propagators in momentum space and then use the
integral representation of the Sinc function to get the final result.
We use the former approach since it involves less algebra.

The Sinc function representation of the fermion field propagator can
be written
\bea
S_{Fh}(p,\Lambda)&=&\frac{hM^2}{32\pi^2}\sum\limits^{+\infty}_{k_1=-\infty}
  \frac{e^{k_1h}}{(\alpha(k_1))^3}\times\nonumber\\
&& \quad \int d^4x\left[ 2m\alpha(k_1)
  -iM^2\gamma^{\mu}x_{\mu}\right] \times\nonumber\\ 
&& \quad \exp{\left(
  -e^{k_1h}\frac{m^2}{M^2}-\frac{x^2M^2}{4\alpha(k_1)}-ipx
  \right).} 
\eea

After taking the Gaussian integral and simplifying we obtain:

\bea
S_{Fh}(p,\Lambda)=\frac{h}{M^2}\sum\limits^{+\infty}_{k_1=-\infty}\exp{\left ( k_1h-e^{k_1h}\frac{m^2}{M^2} \right )}\times\nonumber\\
(m-\gamma^\mu p_\mu)\exp{\left (-\frac{p^2}{M^2}\alpha(k_1) \right )}.
\label{fp_mt}
\eea
Applying the same procedure to the scalar propagator \eq{sc_sinc_coord}  gives
\bea
G_h(p,\Lambda)=\frac{h}{M^2}\sum\limits^{+\infty}_{k_1=-\infty}\exp{\left (k_1h - e^{k_1h}\frac{m^2}{M^2} \right) }\times\nonumber\\
\exp{\left ( -\frac{p^2}{M^2}\alpha(k_1) \right )}.
\label{sc_sinc_mt}
\eea
Finally, the general photon propagator \eq{phot_sinc_gen} and Feynman
gauge photon propagator \eq{phot_sinc_simpl} become
\bea
D^{\mu\nu}_h(p,\Lambda)&=&\frac{h^2}{M^2}\sum\limits_{k_1,k_2}\frac{\exp{
\left( (k_1+k_2)h-e^{k_2h}\frac{m_\nu^2}{M^2} \right
    )}}{\alpha(k_1,k_2)}\times\nonumber\\  
&& \left [\delta^{\mu\nu}\frac{\xi}{2}+(p^2\delta^{\mu\nu}-p^\mu
  p^\nu)\frac{\alpha(k_1,k_2)}{M^2}  \right ] \times \nonumber \\
&& \exp{\left (  -\frac{p^2}{M^2}\alpha(k_1,k_2)\right)} \label{phot_gen_mt}
\eea
and
\bea
\label{phot_simpl_mt}
D^{\mu\nu}_h(p,\Lambda)&=&\delta^{\mu\nu}\frac{h}{M^2}
  \sum\limits_{k_1=-\infty}^{+\infty}\exp{\
  \left ( k_1h-e^{k_1h}\frac{m_v^2}{M^2} \right ) }\times\nonumber\\
&& \quad \exp{\left ( -\frac{p^2}{M^2}\alpha(k_1)\right )}.
\eea

All the machinery necessary for us to start evaluating Feynman
diagrams is now in place.  The simplicity of the propagator in the
Feynman gauge causes us to favor this form of the photon
propagator \eq{phot_simpl_mt} over the more general form
\eq{phot_gen_mt}. The disadvantage of this approach is that
\eq{phot_simpl_mt} is not explicitly transverse, unlike when calculating in
the Landau gauge, $\xi=0$, which automatically ensures that all
computations are consistent with current conservation. Consequently,
using the Feynman gauge requires us take extra steps to make sure
that, despite approximations, the current remains conserved. We will address this issue in detail
when evaluating representative diagrams. However, before calculating
the Feynman graphs we need to determine the error introduced by Sinc
expanding the propagators.

\subsection{Numerical Evaluation}

Since the momentum space propagators were obtained by manipulating
coordinate space versions of the same propagators the error in the
Sinc function approximation to the momentum space propagator is very
similar to that of the position space version. This can be verified by
direct calculation. We work in momentum
space in all further calculations.

It is convenient to start by comparing the approximate version
(\ref{fp_mt}) of the fermion field propagator with the exact propagator
\be
S_F(p)=\frac{m-\gamma^\mu p_\mu}{p^2+m^2}.
\label{fp}
\ee
Both (\ref{fp_mt}) and (\ref{fp}) contain the matrix factor
$m-\gamma^{\mu}p_{\mu}$, which does not depend on the summation
parameter $k_1$ and thus cannot affect the accuracy of the
approximation. This allows us to define a function $\Theta(p,h)$,
which estimates the error associated with the Sinc
expansion of the propagator.
\bea
\Theta(p,h)&=&\frac{h}{M^2}\sum\limits^{+\infty}_{k_1=-\infty} \left[
  \exp{\left (k_1h - e^{k_1h}\frac{m^2}{M^2} \right) }\right. 
  \times\nonumber\\
&& \left. \exp{\left ( -\frac{p^2}{M^2}\alpha(k_1) \right )} \right] 
   -\frac{1}{p^2+m^2}. 
\eea
It is easy to show that the same function can be applied to calculate
the accuracy of the approximations of the scalar field propagator
\eq{sc_sinc_mt} and the photon field propagator \eq{phot_simpl_mt}.

\begin{figure}[tbp]
\begin{center}
\begin{tabular}{c}
\includegraphics[height =8.5cm, angle = -90]{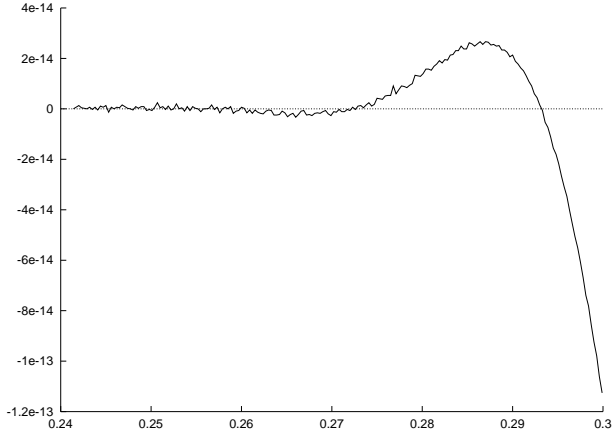}
\end{tabular}
\end{center}
\caption[fig1]{ $\Theta(p,h)$ as a function of $h$ is shown for
  $m=0.511,M=0.511,p=0.75$ and $\Lambda\rightarrow\infty$. Computed
  using 64 bit precision arithmetic.
\label{hdep_fp_double}}
\end{figure}

\Fig{hdep_fp_double} shows $\Theta(p,h)$ as a function of $h$. The
results presented in this plot were obtained using 64 bit precision
arithmetic. From the graph, it is obvious that accuracy better than $1$
part in $10^{-14}$ can be achieved by taking the parameter $h$ smaller
than $0.28$. It is also clear that the approximation stops improving
after $h$ gets smaller than $0.25$. At this point value of
$\Theta(p,h)$ becomes smaller than the numerical noise produced due to
the finite precision of the computation.

\begin{figure}[tbp]
\begin{center}
\begin{tabular}{c}
\includegraphics[height =8.5cm, angle = -90]{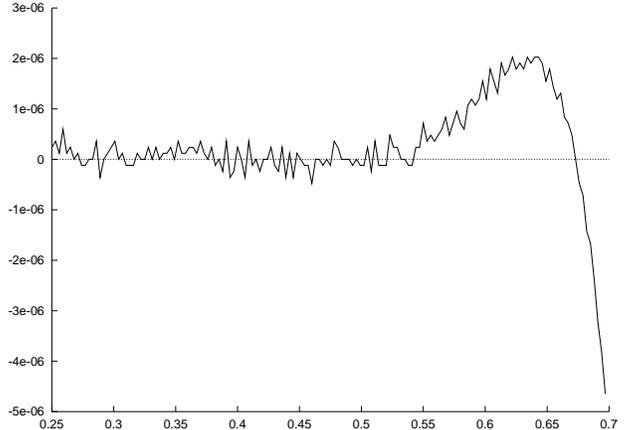}
\end{tabular}
\end{center}
\caption[fig2]{$\Theta(p,h)$ as a function of $h$ is shown for
  $m=0.511,M=0.511,p=0.75$ and $\Lambda\rightarrow\infty$. Computed
  using 32 bit precision arithmetic.
\label{hdep_fp_float}}
\end{figure}

To make sure that the above explanation is valid and that no
other factors limit the accuracy of the approximation, $\Theta(p,h)$
was recomputed using 32 bit precision arithmetic.  The results of this
calculation are presented in \fig{hdep_fp_float}. The smallest error
in this case is around $1$ part in $10^{-6}$, which corresponds to
$h=0.5$. The behavior of $\Theta(p,h)$ is dominated by numerical noise
when $h$ is smaller than $0.5$.

The fact that the value of $h$ at which the approximation breaks
down increases by a factor of 2 as the arithmetic precision of the
computation decreases by the same factor is a consequence of the the
following result, derived in \cite{EastherET1999c}:
\be
|\Theta(p,h)| \le Ce^{-d/h}.
\label{Th_pr}
\ee
The error associated with calculations performed at finite precision
is proportional to $10^{-N}$, where $N$ represents the number of
significant digits. By comparing these two relationships, we conclude
that increasing $N$ by some factor $A$ should expand the range of $h$
in which calculations can be performed with adequate precision to
include values up to $\frac{h}{A}$.

The speculation presented above, \eq{Th_pr} can be verified directly.
Trivial manipulation produces the following result,
\be
\frac{1}{|\log{|\Theta(p,h)|}|}\le \left | \frac{h}{hb-d} \right | \longrightarrow \frac{h}{d}\; \mbox{as} \;h \rightarrow 0,
\ee
where $b$ is a constant that is related to $C$.

\begin{figure}[tbp]
\begin{center}
\begin{tabular}{c}
\includegraphics[height =8.5cm, angle = -90]{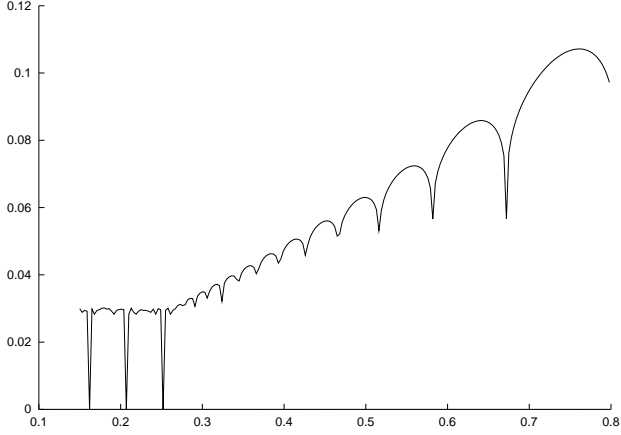}
\end{tabular}
\end{center}
\caption[fig3] {$ 1/|\log{|\Theta(p,h)|}|$ as a function of $h$ is shown for $m=0.511,M=0.511,p=0.2$ and $\Lambda\rightarrow\infty$ .
\label{hdep_fp_log}}
\end{figure}

\begin{figure}[tbp]
\begin{center}
\begin{tabular}{c}
\includegraphics[height =8.5cm, angle = -90]{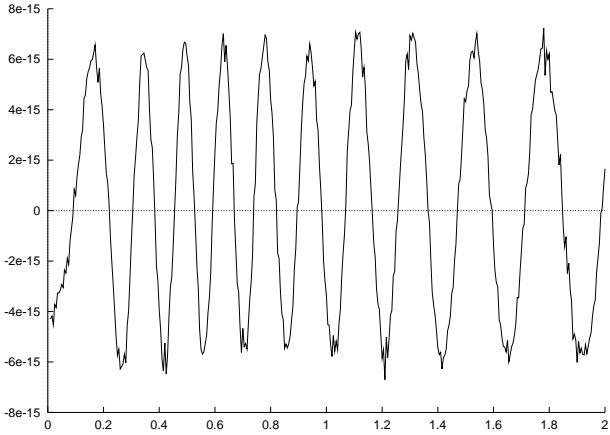}
\end{tabular}
\end{center}
\caption[fig4]{Relative error of the approximation is shown by
  plotting $(p^2+m^2)\Theta(p,h)$ as a function of $p$ for
  $m=0.511,M=0.511,h=0.275$ and $\Lambda\rightarrow\infty$.
\label{pdep_fp}}
\end{figure}

\Fig{hdep_fp_log} shows that for sufficiently small values of $h$ all
maxima of the function $1/|\log{|\Theta(p,h)|}|$ fall on a
straight line. This graph can also be used to determine the smallest
value of the parameter $h$ at which the accuracy of the approximation
is not limited by the numerical precision of the calculation.

Finally, it is necessary to make sure that the approximation holds for
any momentum. A plot of $(p^2+m^2) \Theta(p,h)$ is shown in
\fig{pdep_fp}. Here we see that by setting $h$ to be equal to $0.275$ it
is possible to achieve a relative accuracy of the order of $1$ part in
$10^{-14}$ for a wide range of different momenta. Further, rapid
oscillations of the error allow us to take integrals over momentum
without a significant loss of precision.

\section{Evaluating Feynman Diagrams}

\subsection {The Electron Vertex Function}
We use the Sinc function representation to compute one-loop diagrams
in QED, in accordance with the {\it Sinc function Feynman rules},
given in \cite{EastherET1999c}.  Since results for these diagrams are
well known, we can use them as a first test of the approximation.
Furthermore, it will be necessary to have expressions for these graphs
to perform higher order calculations.

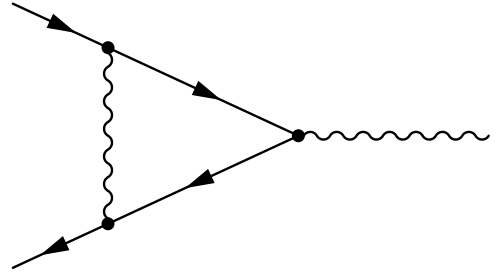
\begin{figure}
\begin{center}
\begin{fmffile}{vertex}
\begin{fmfgraph*}(200,100)
  \fmfleft{i1,i2}
  \fmfright{o1}
  \fmf{fermion,label=$p$,tension=2}{i2,v1}
  \fmf{fermion,label=$k$}{v1,v2}
  \fmf{fermion,label=$k'$}{v2,v3}
  \fmf{fermion,label=$p'$,tension=2}{v3,i1}
  \fmf{photon,tension=2,label=$q$}{v2,o1}
  \fmffreeze
  \fmf{photon,label=$p-k$}{v1,v3}
  \fmfdotn{v}{3}
\end{fmfgraph*}
\end{fmffile}
\end{center}
\caption[fig5]{\label{vertex} The Electron Vertex Diagram}
\end{figure}

We start by evaluating the electron vertex.  This diagram, shown in
\fig{vertex}, contributes $\alpha/(2\pi)$ to the anomalous magnetic
moment of the electron. In the following, we reproduce this result
using the Sinc
function representation.\footnote{In
  this paper we focus our attention on numerical calculations of
  graphs which have been evaluated analytically, to ensure we have
  benchmarks for assessing the accuracy of the Sinc function
  representation. The analytic results quoted in this section can be
  found in any thorough QED text, such as Peskin and Schroeder
  \cite{PeskinBK1}.}

The expression for the leading correction to vertex function is
\bea
\bar{u}(p')\delta\Gamma^{\mu}(p,p')u(p)=\bar{u}(p')\bigg
{[}\int\frac{d^4_Ek}{(2\pi)^4}D^{\rho\nu}(p-k)\times\nonumber\\ 
(-e\gamma_{\nu})S_F(k')\gamma^\mu S_F(k)(-e\gamma_{\rho}) \bigg{]}u(p).
\eea

After substituting our expressions for the propagators we find:
\bea
&&\bar{u}(p')\delta\Gamma^{\mu}(p,p')u(p)= \nonumber\\
&& \quad\bar{u}(p')\bigg {[}\frac{e^2h^3}{M^6(2\pi)^4}
  \sum\limits_{k_1,k_2,k_3}\int d^4_Ek\times\nonumber \\ 
&& \quad \exp{\left \{(k_1+k_2+k_3)h-e^{k_1h}\frac{m_\nu^2}{M^2}-
  \right.}\nonumber\\ 
&& \qquad \left. 
\left(e^{k_2h}+e^{k_3h}\right )\frac{m_e^2}{M^2}   \right\}
 \times \nonumber \\ 
  && \quad 
\gamma_\rho(m_e-(k^{\sigma}+q^{\sigma})\gamma_\sigma)) \gamma^\mu
 (m_e-k^\lambda\gamma_\lambda)\gamma^\rho\times\nonumber\\
&& \quad \exp{\left
    \{-\frac{(p-k)^2}{M^2}\alpha(k_1)-\frac{(k+q)^2}{M^2}\alpha(k_2)-
\right. }\nonumber\\ 
&& \qquad \left. 
 \frac{k^2}{M^2}\alpha(k_3) \right\} \bigg{]}u(p).  
\label{horror}
\eea

We eliminate gamma matrices with repeated indices, take the Gaussian
integral over $k$, use the Dirac equation and
\bea
\gamma^\mu p_\mu u(p)=-m_eu(p),\\
\bar{u}(p')\gamma^\mu p'_\mu=-\bar{u}(p)m_e
\eea
to simplify the result.  Performing all the steps described above and
using the Gordon identity,
\be
\bar{u}(p')\gamma^\mu(p)=\bar{u}(p')\left[
  \frac{p'^\mu+p^\mu}{2m_e}+\frac{i\sigma^{\mu\nu}q_\nu}{2m_e}\right] 
u(p) ,
\ee
to eliminate terms proportional $(p^\mu+p'^\mu)$ we obtain the
following expression for the second order correction to the electron
magnetic moment:
\bea
\frac{g-2}{2}&=&\frac{\alpha}{\pi}\frac{m_e^2h^3}{M^2}\sum\frac{1}{\Delta^2}
\frac{\alpha(k_1)}{\Delta}\left(
  1-\frac{\alpha(k_1)}{\Delta} \right )\times\nonumber\\
&&\exp{\left\{
  (k_1+k_2+k_3)h\right\}}\times\nonumber\\
&&\exp{\left\{
  -e^{k_1 h}\frac{m_v^2}{M^2}-\left(e^{k_2 h}+e^{k_2 h}
  \right)\frac{m_e^2}{M^2}\right\}}\times\nonumber\\  
&&\exp{\left\{ 
  \frac{m_e^2}{M^2\Delta}(\alpha(k_2)+\alpha(k_3))\alpha(k_1)  \right \}}.
\eea
Here $\Delta=\alpha(k_1)+\alpha(k_2)+\alpha(k_3)$. 

The results of numerically evaluating this sum for different values of
$h$ and $m_e=M=0.511$ are presented in Table~\ref{vertex_tbl}.  As $h$
is lowered the error decreases to a few parts in $10^{13}$. However,
when $h \lesssim 0.3$ the accuracy decreases. This is due to the use
of fixed precision (64 bit) arithmetic, and is not intrinsic to the
Sinc function representation itself.

Computationally, the triple sum is simple to evaluate, and for typical
values of $h$ it can be computed on a desktop computer in less than a
second.  In general, the time taken to evaluate a given diagram is a
function of the number of internal propagators. The sums generated
from QED diagrams are not significantly more complex than those
derived from scalar diagrams with the same number of propagators. More
detailed analysis of the relationships between the form of the diagram
and the complexity of its Sinc function representation, as well as
discussions of efficient summation algorithms, are given in
\cite{EastherET1999c,EastherET1999e}.

\begin{table}
\caption{\label{vertex_tbl}Correction to the magnetic moment for
  different values of $h$.} 
\begin{center}
\begin{tabular}{|c|c|c|}\hline
h & Correction$\times\frac{\alpha}{\pi}$ & 
   Error$\times\frac{\alpha}{\pi}$ \\ \hline
0.2 & 0.4999999999949221 & $ 5.1 \times 10^{-12}$\\ \hline
0.4 & 0.5000000000008583 & $ -8.6 \times 10^{-13}$\\ \hline
0.6 & 0.500000008001 & $ -8.0 \times 10^{-9}$\\ \hline
0.8 & 0.499999900285 & $  9.97 \times 10^{-8}$\\ \hline
1.0 & 0.499998 & $  2.3 \times 10^{-6}$\\ \hline
1.2 & 0.49977 & $ 0.00023 $\\ \hline
1.4 & 0.4987 & $ 0.0012 $\\ \hline
\end{tabular}
\end{center}
\end{table}

\subsection{Electron Self-Energy}

\begin{figure}
\begin{center}
\begin{fmffile}{celfenergy}
\begin{fmfgraph*}(150,50)
  \fmfleft{i1}
  \fmfright{o1}
  \fmf{fermion,label=$p$,tension=1.5}{i1,v1}
  \fmf{fermion,label=$p-k$}{v1,v2}
  \fmf{fermion,label=$p$,tension=1.5}{v2,o1}
  \fmffreeze
  \fmf{photon,right,label=$k$}{v2,v1}
  \fmfdotn{v}{2}
\end{fmfgraph*}
\end{fmffile}
\end{center}
\caption[fig6]{\label{celfenergy} Electron self-energy diagram.}
\end{figure}
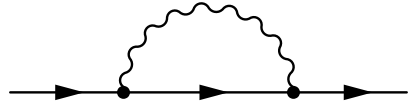

The second order electron self-energy diagram is shown in
\fig{celfenergy}. The quantity, $\Sigma(p)$, which represents a shift
of the electron mass, can be written as follows.

\be
\Sigma(p)=-\int\frac{d^4k}{(2\pi)^4}(-e\gamma_\mu)S(p-k)(-e\gamma_\nu)D^{\mu\nu}(k)
\ee

Substitution of the Sinc function representations of the propagators
transforms this to
\bea
\Sigma(p)&=&-\frac{h^2e^2}{(2\pi)^4M^4}\times\nonumber\\
&&\sum\limits_{k_1,k_2}\exp{\left
    \{(k_1+k_2)h-e^{k_1h}\frac{m_e^2}{M^2}
    -e^{k_2h}\frac{m_\nu^2}{M^2} \right \}}\nonumber\\ 
&&\int d^4k(\gamma_\mu(m_e-\gamma^\sigma p_\sigma +\gamma^\sigma
k_\sigma) \gamma^\mu)\times\nonumber\\ 
&&\exp{\left
    \{-\frac{(p-k)^2}{M^2}\alpha(k_1)-\frac{k^2}{M^2}\alpha(k_2)  \right \}}.
\eea
After taking the Gaussian integral, eliminating gamma matrices and
simplifying, we obtain
\bea
\Sigma(p)&=&\left(\frac{\alpha}{\pi}\right )\frac{h^2}{2}
 \sum\limits_{k_1,k_2}
   \frac{2m_e+p^\sigma\gamma_\sigma\left(
       1-\frac{\alpha(k_1)}{\Delta}\right)}
   {\Delta^2}\times\nonumber\\ 
&&\exp{\left\{ (k_1+k_2)h-e^{k_1h}\frac{m_e^2}{M^2}-e^{k_2h}\frac{m_\nu^2}{M^2}
  \right\}}  \times\nonumber\\ 
&&\exp{\left\{-\frac{\alpha(k_1)\alpha(k_2)p^2}{M^2\Delta} \right\}}.
\label{SE_1}
\eea
Here $\Delta=\alpha(k_1)+\alpha(k_2)$. 

The quantity above diverges as $\Lambda$ goes to infinity.  In order
to renormalize it,
 we follow standard procedures. First we isolate the
matrix dependence by the decomposition,
\be
\Sigma(p)=A(p^2)+p^\sigma\gamma_\sigma B(p^2),
\ee
where $A(p^2)$ and $B(p^2)$ are easily determined from equation \eq{SE_1}.
The finite quantity $\Sigma_R(p)$ is formed by subtraction of the
divergent parts of $\Sigma(p)$.
\be
\Sigma_R(p)=(A(p^2)-A(0))+p^\sigma\gamma_\sigma (B(p^2)-B(0)).
\label{SE_ren}
\ee
Combining expressions \eq{SE_ren} and  \eq{SE_1} we obtain:
\bea
\Sigma_R(p)&=&\left(\frac{\alpha}{\pi}\right)
  \frac{h^2}{2}\sum\limits_{k_1,k_2} 
\frac{2m_e+p^\sigma\gamma_\sigma\left
    (1-\frac{\alpha(k_1)}{\Delta}\right)}{\Delta^2}\times\nonumber\\ 
&&\exp{\left
    \{(k_1+k_2)h-e^{k_1h}\frac{m_e^2}{M^2}-e^{k_2h}\frac{m_\nu^2}{M^2}
  \right \}} \times\nonumber\\ 
&&\left [ \exp{\left\{-\frac{\alpha(k_1)\alpha(k_2)p^2}{M^2\Delta}
    \right\}}-1\right ]. 
\label{SE_2}
\eea

\begin{figure}[tbp]
\begin{center}
\begin{tabular}{c}
\includegraphics[height =8.5cm, angle = -90]{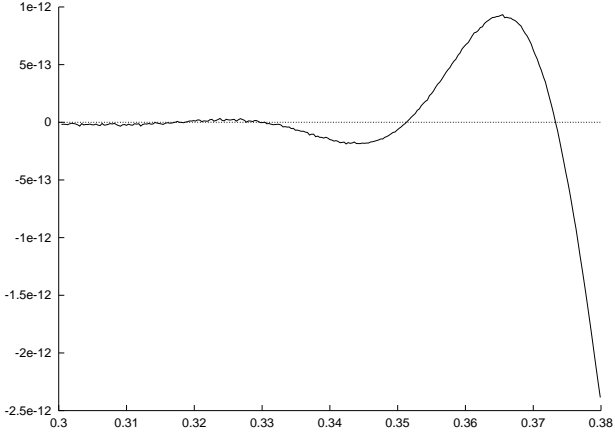}
\end{tabular}
\end{center}
\caption[fig7]{The absolute error in $\Sigma_R(p)$  as a function of
  $h$ is shown for $m=0.511,M=0.511,p=0.75$ and 
  $\Lambda\rightarrow\infty$.\label{hdep_se_double}}
\end{figure}

The difference between the values of $\Sigma_R(p)$ obtained from
\eq{SE_2} and the exact propagator correction to this order,
\bea
\Sigma_R(p)&=&\left(\frac{\alpha}{\pi}\right) 
  \frac{h^2}{2}\int\limits_{0}^{1}(2m_e+xp^\sigma\gamma_\sigma)
 \times\nonumber\\
&&\log{\left (\frac{(1-x)m_e^2+xm_\nu^2}
  {(1-x)m_e^2+xm_\nu^2+x(1-x)p^2} \right)},\;\;\;\;
\eea
is shown as a function of $h$ in \fig{hdep_se_double}. By setting $h$
to be equal to $0.31$ it is possible to achieve accuracy better than
$1$ part in $10^{14}$ in this calculation.

\subsection{Charge Renormalization}

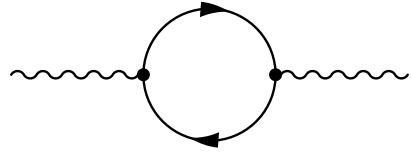
\begin{figure}
\begin{center}
\begin{fmffile}{elloop}
\begin{fmfgraph*}(150,50)
  \fmfleft{i1}
  \fmfright{o1}
  \fmf{photon,label=$p$,tension=2}{i1,v1}
  \fmf{photon,label=$p$,tension=2}{v2,o1}
  \fmf{fermion,left,label=$p+k$}{v1,v2}
  \fmf{fermion,left,label=$k$}{v2,v1}
  \fmfdotn{v}{2}
\end{fmfgraph*}
\end{fmffile}
\end{center}
\caption[fig8]{\label{ChRen} Electron loop diagram.}
\end{figure}

To second order in perturbation theory, the electron loop diagram
$\Pi^{\mu\nu}(p)$, shown in \fig{ChRen}, is the only relevant graph.
$\Pi^{\mu\nu}(p)$ has the familiar form:
\be
\Pi^{\mu\nu}(p)=e^2\int\frac{d^4k}{(2\pi)^4}tr\left[
  \gamma^{\mu}S(k)\gamma^{\nu}S(p+k) \right]. 
\ee
After substituting Sinc expanded representations for $S(k)$ and
$S(p+k)$ and taking the trace, we obtain
\bea
\Pi^{\mu\nu}(p)&=&\frac{e^2h^2}{4M^4\pi^4}\sum\limits_{k_1,k_2}    
  \nonumber\\  
&&  \exp{\left\{(k_1+k_2)h-\left(e^{k_1h}+e^{k_2h} \right)
  \frac{m_e^2}{M^2} \right\}}\times\nonumber\\
&&\int d^4k[p^\mu k^\nu+2k^\mu k^\nu+k^\mu p^\nu- \nonumber\\ 
&&  \delta^{\mu\nu}(p\cdot k+k^2+m_e^2)]\times\nonumber\\
&&\exp{\left\{-\frac{k^2}{M^2}\alpha(k_2)-\frac{(p+k)^2}{M^2}\alpha(k_1)
   \right\}}.
\label{CR_1}
\eea

By now it is not surprising that the integration over internal loop
momentum $k$ can be easily performed analytically.  As we mentioned
earlier, in order to use this (or any approximation) we must be sure
that current conservation is respected.  We achieve this by
multiplying $\Pi^{\mu\nu}$ by the transverse projection operator
\be
\left(\frac{p^\lambda p^\sigma}{p^2}-\delta^{\lambda\sigma} \right).
\label{proj}
\ee
After taking the integral in (\ref{CR_1}) and projecting out the 
transverse part with (\ref{proj}) we obtain 
\bea
&&\Pi^{\mu\nu}(p)=\frac{\alpha}{\pi}h^2\sum\limits_{k_1,k_2}\frac{1}
 {\Delta^2}\exp{\left   
    \{-\frac{\alpha(k_1)\alpha(k_2)p^2}{\Delta M^2} \right
  \}}\times\nonumber\\ 
&&\exp{\left\{(k_1+k_2)h-\left(e^{k_1h}+e^{k_2h}
    \right)\frac{m_e^2}{M^2} \right\}}\times\nonumber\\ 
&&\left
  (\frac{\alpha(k_1)\alpha(k_2)p^2}{\Delta^2}-m_e^2-\frac{M^2}{\Delta}
\right )\left (\frac{p^\mu p^\nu}{p^2}-\delta^{\mu\nu} \right).
\label{CR_2}
\eea
It is convenient to make the usual definition:
\be
\Pi^{\mu\nu}(p)=\Pi(p^2)\left (\frac{p^\mu p^\nu}{p^2}-\delta^{\mu\nu} \right).
\ee
The quantity $\Pi(p^2)$ diverges as $\Lambda\longrightarrow\infty$. We
can renormalize it by subtracting out divergent parts, which are
contained in the first two terms of the Taylor series
\be
\Pi_R(p^2)=\Pi(p^2)-\Pi(0)-p^2\left . \frac{d \Pi(p^2)}{dp^2}\right |_{p^2=0}.
\ee
Combining the definitions above with equation \eq{CR_2}, we derive the
final result
\bea
\Pi_R(p^2) &=&
\frac{\alpha}{\pi}h^2\sum\limits_{k_1,k_2}\frac{1}{\Delta^2}
\nonumber\\
&&\exp{
  \left\{(k_1+k_2)h-\left(e^{k_1h}+e^{k_2h}\right)\frac{m_e^2}{M^2} 
    \right\}} \times\nonumber\\ 
&&\left[\left( \frac{\alpha(k_1)\alpha(k_2)p^2}{\Delta^2}- 
  m_e^2-\frac{M^2}{\Delta}  
   \right)\right.\times  \nonumber\\
&& \left.\exp{\left\{ -\frac{\alpha(k_1)\alpha(k_2)p^2}{\Delta M^2} \right\}
  } \right]
+  m_e^2+\nonumber\\ 
&&\left.\frac{M^2}{\Delta}-\frac{\alpha(k_1)\alpha(k_2)} 
 {\Delta}p^2\left (\frac{2}{\Delta}+\frac{m_e^2}{M^2} \right)\right ]. 
\eea

\begin{figure}[tbp]
\begin{center}
\begin{tabular}{c}
\includegraphics[height =8.5cm, angle = -90]{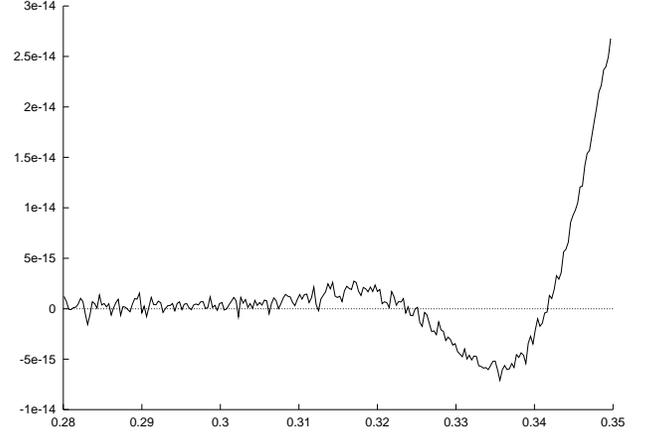}
\end{tabular}
\end{center}
\caption[fig9]{Absolute error in $\Pi_R(p^2)$ is shown as a function
  of $h$ for $m_e=0.511, M=0.511, p=0.75$. 
\label{hdep_cr_double}}
\end{figure}

As before we check the accuracy of the above equation by 
comparing it to the exact result,
\bea
\Pi_R(p^2)&=& 2\frac{\alpha}{\pi}\int\limits_0^1dx\:x(1-x)p^2 \times
\nonumber\\
 && \qquad  \log{\left (\frac{m_e^2}{m_e^2+x(1-x)p^2} \right)}.
\eea
The result of this comparison is presented in \fig{hdep_cr_double}.
The error of the approximation stops decreasing after it reaches a
value close to a $1$ part in $10^{15}$ at $h=0.3$.

\section{Conclusions}

In this paper we have extended the Sinc function techniques developed
in \cite{EastherET1999c} to include fermion and vector propagators.
We display approximate forms of scalar, fermion and photon field
propagators derived in both coordinate and momentum space.  The ease
of obtaining high accuracy with these approximations is demonstrated
by performing several numerical comparisons between exact and Sinc
expanded versions of the propagators. These comparisons show that
using 64-bit numerical precision, currently available on all hardware,
it is possible to achieve errors as low as a few parts in $10^{15}$.  On
the other hand, if high precision can be sacrificed for speed, it is
possible to increase the parameter $h$ with a corresponding decrease
in accuracy. However, even the highest accuracy calculations described
in this paper typically take less than a second on standard desktop
computers.

To illustrate the use of the Sinc function representation and lay the
ground for future work we evaluated three one-loop QED Feynman
diagrams. In all three cases it is easy to perform the calculations by
hand, so we were able to evaluate the error in computations based on
Sinc approximated propagators. In general, there are three factors
that limit the accuracy of the calculation. First, some error is
introduced when we use the Sinc representation to approximate the
propagators. It is possible to improve the approximation by decreasing
the value of $h$.  Second, truncation of the sums, which have to be
evaluated in order to compute Feynman graphs using the Sinc expansion,
produces another contribution to the total error.  Finally, the finite
numerical precision of the computation is the third source of error.

The three sources of error are closely related. For example, an
attempt to achieve higher accuracy by decreasing the value of $h$
without simultaneously increasing the range of the summation would
result in a loss of precision. To minimize the error, all parameters
must be set to optimal values. While it is easy to manipulate $h$ and
the range of the summation, increasing numerical precision can be
difficult since we often have to evaluate terms which involve
subtracting two almost equal quantities.

The results in this paper provide further confirmation that the Sinc
function representation can be used in high precision calculations
with the expenditure of surprisingly small amounts of computing time.
Higher order calculations will require significantly more time than
those presented in this paper, but results to date indicate that Sinc
techniques can still be orders of magnitude faster than Monte Carlo
methods when high accuracy is required.  Finally, the methods we have
presented here can, in principle, be automated. In future work we plan
to focus on this task in order to facilitate the calculation of higher
order perturbative graphs in arbitrary theories.

\section*{Acknowledgments}

We thank Pinar Emirda\u{g} for very useful
discussions. This work was supported by the United States Department
of Energy, via contract DE-FG0291\-ER40688, Tasks A and D.  The
computational work described in this paper was carried out at the
Theoretical Physics Computing Facility at Brown University.

\end{document}